\documentclass[12pt]{article}

\usepackage[toc,page,title,titletoc,header]{appendix}
\usepackage[T1]{fontenc}
 \usepackage{color}
\usepackage{enumerate}
\usepackage{graphicx}
\usepackage[latin1]{inputenc}
\usepackage{epstopdf}
\usepackage{amsthm}
\usepackage{amsmath}
\usepackage{shapepar}
\usepackage{times}

\usepackage{amssymb, amsbsy, amsthm}
\makeatletter

\newtheorem{Theorem}{Theorem}
\newtheorem{Lemma}{Lemma}

\newenvironment{sciabstract}{%
\begin{quote} \bf}
{\end{quote}}

\title{Certifying randomness in quantum state collapse}

\author
{Liang-Liang Sun,$^{1}$ Xingjian Zhang,$^{2}$ Xiang-Zhou,$^{1}$ Zheng-Da Li,$^{3,4}$ \\
Xiongfeng Ma,$^{2}$ Jingyun Fan, $^{3,4\dag}$ Sixia Yu,$^{1\ddag}$\\
\\
\normalsize{$^{1}$Hefei National Laboratory for Physical Sciences at the Microscale and  }\\
\normalsize{Department of Modern Physics, University of Science and Technology of China,}\\
\normalsize{  Hefei, Anhui 230026, China}\\
\normalsize{$^{2}$Center for Quantum Information,
Institute for Interdisciplinary }\\
\normalsize{Information Sciences, Tsinghua University, Beijing, China }\\
\normalsize{$^{3}$Shenzhen Institute for Quantum Science and Engineering and Department of Physics,}\\
\normalsize{ Southern University of Science and Technology, Shenzhen, 518055, China}\\
\normalsize{$^{4}$ Guangdong Provincial Key Laboratory of Quantum Science and Engineering, }\\
\normalsize{Southern University of Science and Technology, Shenzhen, 518055, China,}\\
\\
\normalsize{$^\dag$E-mail: fanjy@sustech.edu.cn}\\
\normalsize{$^\ddag$E-mail:yusixia@ustc.edu.cn}
}

\date{}

\begin{document}


\baselineskip24pt


\maketitle


\begin{sciabstract}
  The unpredictable process of state collapse caused by quantum measurements makes the generation of quantum randomness possible. In this paper, we explore the quantitative connection between the randomness generation and the state collapse and provide a randomness verification protocol under the assumptions: (I) independence between the source and the measurement devices and (II) the L\"{u}ders'  rule for collapsing state. Without involving heavy mathematical machinery, the amount of generated quantum randomness can be directly estimated with the disturbance effect originating from the state collapse. In the protocol, we can employ  measurements that are  trusted to be non-malicious but not necessarily be characterized. Equipped with trusted and characterized  projection measurements, we can further optimize the randomness generation performance. Our protocol also shows a high efficiency and yields a higher randomness generation rate than the one based on uncertainty relation. We expect our results to provide new insights for understanding and generating quantum randomness.
\end{sciabstract}


\section{Introduction}
Randomness is ubiquitous in modern society. In particular, it plays an indispensable role in cryptography. Such a resource is absent within the deterministic Newtonian physics. On the contrary, there is an ample supply of intrinsic randomness in a quantum world. Many quantum properties, such as nonlocality, uncertainty principle, and contextuality, can ensure the presentation of quantum randomness and have been harnessed to devising quantum randomness generators (QRGs)~\cite{Ekert1991Quantum, PhysRevLett.95.010503, colbeck2011quantum, PhysRevLett.108.100402, 2009Masanes, 2010Random, 2018Device, PhysRevA.90.052327, 2014furrer, PhysRevLett.109.100502, PhysRevX.6.011020, PhysRevA.86.062109, 2018Corrigendum, PhysRevLett.119.240501}.%
These properties deal with various scenarios where specific functioning of quantum devices are required, for example,  the system dimension~\cite{PhysRevA.84.034301, PhysRevA.85.052308, PhysRevLett.114.150501, 2016guo}, indistinguishability of non-orthogonal quantum states~\cite{PhysRevApplied.7.054018}, and energy of system~\cite{VanHimbeeck2017, 2020rusca, Tebyanian2021}. These requirements are the assumptions or the expense of the revelent QRNGs.
Among all quantum randomness generation schemes, device-independent (DI) protocols enjoy the highest security with almost the only assumption on the correctness of quantum physics~\cite{colbeck2011quantum}, however, which is  extremely challenging in the experiment and could achieve only very low rates of randomness generation.  The other extreme are fully trusted quantum randomness generators, where more randomness is easily extracted at the expense of fully trusting the inner working of physical devices, which is most unscure but generate random number at a high rate. The semi-DI protocols are the compromise between security and rate of  generation, in which the central problem is finding protocols of operationally simple, high randomness generating rate,  and  fewer security assumptions.

In this paper, we exploit the common knowledge on QRNG: no matter what quantum property it involves, state collapse induced by measurements, which is the solely unpredictable process in quantum theory, has to be present.
We explore the possibility of directly verifying quantum randomness with state collapse. By using the disturbance effect accessible in a sequence of incompatible measurements, we provide a confirmative answer to the question: we establish a quantitative connection between randomness generation, state collapse, and the disturbance effect. We employed a prepare-and-measure QRNG scenario to demonstrate this connection.
It involves an untrusted source of quantum states and two quantum measurements performed in sequence, which is readily implementable on photonic experimental platforms. With a few reasonable assumptions on the device's functioning, the protocol can employ a general unknown general measurement trusted to be non-malicious  and also allows for optimizing the performance using completed trusted and characterized projection measurements. In various contexts, quantum randomness generated via our protocol can be directly estimated without involving heavy mathematical machinery. Thus, we provide an efficient RNG protocol as well as a quantitative account for the fundamental connection between the key concepts, namely, quantum randomness and state collapse.

The rest of the paper is structured as follows. In section I, we briefly review measures of quantum randomness. In section II, we introduce the set-up of our protocol. In section III, we establish the connection between disturbance and quantum randomness against a  classical adversary. We show that our the performance of our QRNG protocol can be optimized when more information about measurements is at hand. In section IV, we use the protocol to estimate the quantum randomness against classical and the quantum adversaries in the asymptotic limit of an infinite data size. In section V, we compare our result with the protocol based on uncertainty relation.

\section{Quantum Randomness Measures}
In information theory, quantum randomness evaluation can be formalized in an adversarial scenario~\cite{randommeasure}.
Consider a user, Alice, and an adversary, Eve, share particles in a joint state $\rho_{AE}$. A local measurement $\mathcal{M}_{A}$ performed on the subsystem of Alice alters the entire state from $\rho_{AE}$ to $\rho'_{AE}$. Because of the presence of Eve's side information, Alice's measurement results are not completely private.
Depending on Alice's measurement and Eve's adversary strategies, different entropic measures may be applied to quantify the amount of private randomness. In a generic single-shot case, one applies the conditional min-entropy as the randomness measure. Generally, Eve may utilise the full knowledge of her system, and the conditional min-entropy is defined as
\begin{equation}\label{Eq:QuantumEve}
\operatorname{H}^{Q}_{min}(A|E)=-\textstyle\inf_{\sigma_{E}}\operatorname{D}_{max}(\rho'_{AE}\|\operatorname{id}_A\otimes \sigma_{E}), 
\end{equation}
where $\operatorname{id}$ denotes the identity operator, $\sigma_{E}$ is a normalized state on Eve's system, and $\operatorname{D}_{max}(\rho\|\sigma)$ is the maximum relative entropy,
\begin{equation}
  \operatorname{D}_{max}(\rho\|\sigma)=\inf\{\lambda \in\mathbb{R}:\rho\leq 2^{\lambda}\sigma\}.\nonumber
\end{equation}
In certain contexts, the potential side information has a classical nature. Operationally, this corresponds to the case where Eve carries out a measurement on her system and use the measurement outcome as her guess. Then, the conditional min-entropy degenerates to the following quantity,
\begin{equation}\label{Eq:ClassicalEve}
\operatorname{H}^{C}_{min}(A|E)=-\textstyle\inf_{\sigma_{E}}\operatorname{D}_{max}(\rho''_{AE}\|\operatorname{id}\otimes \sigma_{E}),
\end{equation}
where $\rho''_{AE}$ is the post-measurement state after Alice's and Eve's local measurements. Depending on whether Eve's side information is characterised by a quantum state or a classical random variable, we call the entropic measures in Eq.~\eqref{Eq:QuantumEve} and~\eqref{Eq:ClassicalEve} as conditioned on a quantum adversary and a classical adversary, respectively.

From the adversarial perspective, quantum randomness is conversely associated with the maximum probability that Eve can correctly guessing the outcomes on Alice's side, which we call the guessing probability. For the case where Alice performs measurement $\mathcal{M}_{A}=\{M_{i}\}$ on a pure state $|\phi\rangle$, the best adversarial strategy for Eve is simply guessing the outcome with the maximum probability, given by
$$\operatorname{G}_{A|\phi}:=\textstyle\max_{i}p(i|A; \phi),$$ where $p(i|A; \phi)=\langle\phi|M_i|\phi\rangle$.
For a mixed state $\rho$, Eve can utilise her side information for a better guess. In the case of a classical adversary, the guessing probability is given by
   \begin{eqnarray*}
\operatorname{G}_{A|\rho}=\textstyle \max_{\{r_{n},\phi_{n}\}} \textstyle\sum_{n} r_{n}\cdot \operatorname{G}_{A|\phi_{n}}.
  \end{eqnarray*}
 where the optimization is taken over all pure state decompositions $\rho=\sum_{n}r_{n}|\phi_{n}\rangle\langle \phi_{n}|$.
The conditional min-entropy in Eq.~\eqref{Eq:ClassicalEve} has the following equivalent expression,
\begin{eqnarray}
 \textstyle\operatorname{H}^{C}_{min}(A|\rho)=-\log \textstyle \operatorname{G}_{A|\rho}.
  \end{eqnarray}

When Alice repeats projective measurements in basis $\{|i\rangle\}$  independently and identically for sufficient times, the conditional entropy Eq.~\eqref{Eq:QuantumEve} asymptotically converges to~\cite{PhysRevA.92.022124, 2018, 2019}
\begin{eqnarray}
  \operatorname{H}^{Q}_{min, asy}(A|E)=S(\rho\|\Delta(\rho)),\label{cr}
  \end{eqnarray}
where $\Delta(\rho):=\sum_{i}|i\rangle\langle i|\rho|i\rangle\langle i|$ and the relative entropy $S(\rho\|\Delta(\rho)):=\operatorname{tr}\rho[\log \rho-\log \Delta(\rho)]$. The conditional Eq.\eqref{Eq:ClassicalEve} asymptotically converges to~\cite{PhysRevA.92.022124, 2018, 2019}
\begin{eqnarray}
\operatorname{H}^{C}_{min, asy}(A|E)=\textstyle\min_{r_{n}, \phi_{n}} \textstyle\sum_{n} r_{n}\cdot \operatorname{H}(\mathbf {p_{\phi_n}}),  \label{cf}
  \end{eqnarray}
 with  $\mathbf{p_{\phi_{n}}}=\{p(i|A, \phi_{n})\}$ and $ \operatorname{H}(\cdot)$ being  the Shannon entropy.

\section{Quantum Randomness Verification based on Disturbance}
Fundamentally, a random number generating measurement must cause state collapse. Consider a simple example of measuring $\sigma_{z}$  respectively on two states, $\{\{\frac{1}{2}, |0\rangle\langle 0|\}; \{\frac{1}{2}, |1\rangle\langle 1|\}\}$ and $\frac{\sqrt{2}}{2}(|0\rangle+|1\rangle)$. The outcome probabilities in both cases are uniform. For the former case, in each run of the measurement, the observable takes a definite value and hence the measurement  neither collapses the measured state  nor produces quantum randomness. In contrast, measuring $\sigma_{z}$ on the state $\frac{\sqrt{2}}{2}(|0\rangle+|1\rangle)$ introduces the maximum state disturbance and produces one bit of quantum randomness in each run. In the following, we generalize the above observation to a more general scenario and relate randomness and disturbance in a randomness generation protocol.

\subsection{Protocol}
\begin{figure}
\centering
\includegraphics[width=0.6\textwidth]{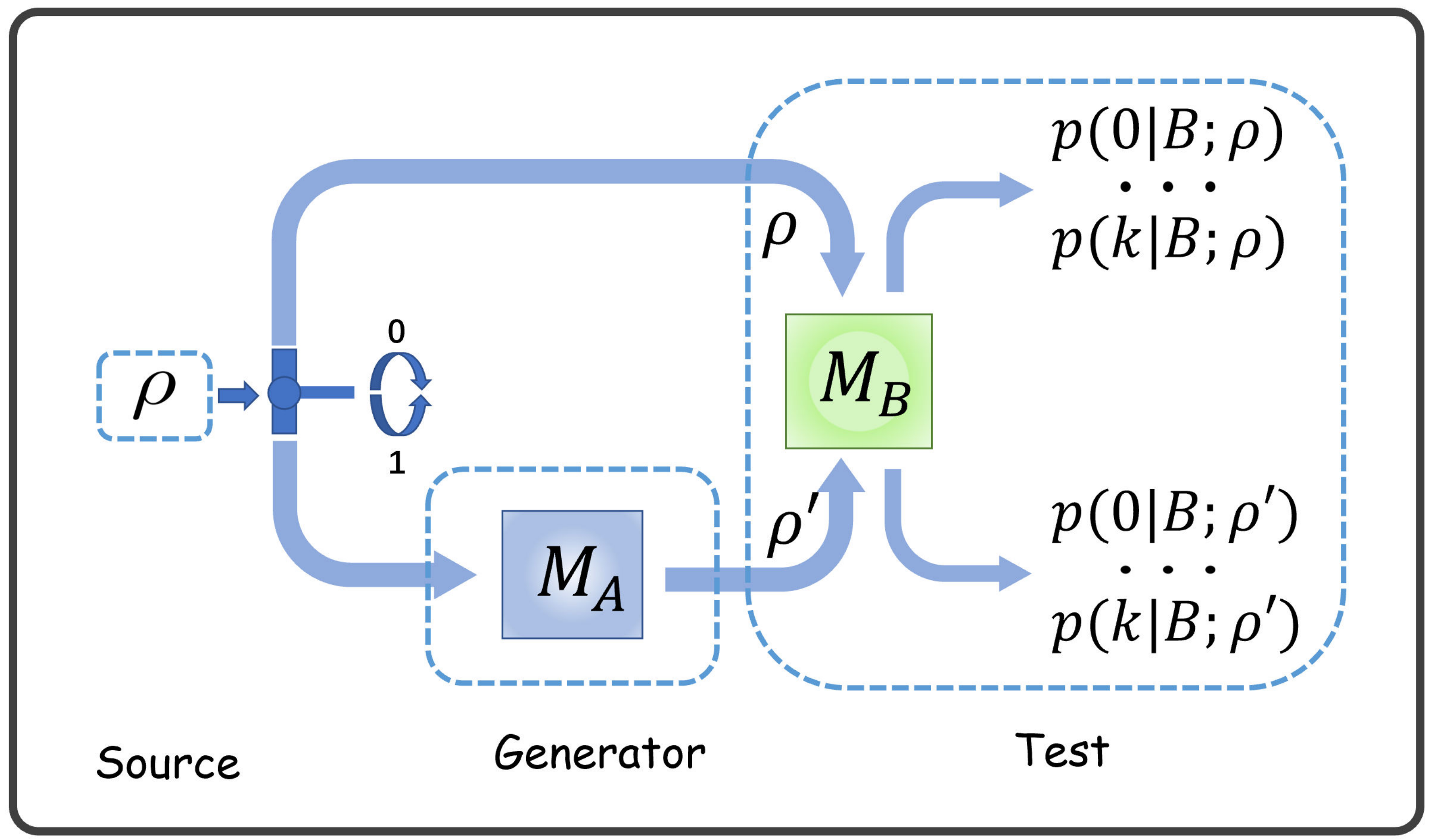}
\caption{QRNG protocol: The protocol consists of a source, a randomness generating measurement ($\mathcal{M}_{A}$), and a randomness test measurement ($\mathcal{M}_{B}$). The source prepares particles in an unknown state $\rho$,  which may be correlated with an adversary's devices. The particles are randomly sent to one of the paths, which may be achieved via a switch controlled by randomness numbers.  In the lower path,  measurement $\mathcal{M}_{A}$ is performed,  which yields raw random numbers, and the input state is transferred into $\rho'$ on average. In the upper path, nothing is done, and the particles are transferred to the following measurement $\mathcal{M}_{B}$. To test the genuine randomness contained in the raw data, a test measurement $\mathcal{M}_{B}$ respectively on $\rho'$ and on $\rho$. By the yielded distributions, the degree of state collapse can be estimated in terms of the disturbance effect of $\mathcal{M}_A$ in $\mathcal{M}_{B}$, and the extractable random number is bounded. }
\label{fig1}
\end{figure}

As shown in Fig.~\ref{fig1}, we consider a prepare-and-measure scenario, which consists of an untrusted state source and two  measurements. The source prepares a state, $\rho$,  a randomness-generating measurement, $\mathcal{M}_A$, and a randomness-testing measurement, $\mathcal{M}_B$, to verify the state collapse. Our protocol consists of five steps:
\begin{itemize}
\item[(1)] Prepare particles in a state, $\rho$.

\item[(2)] With a pre-fixed probability distribution, every  particle randomly undergoes one of the following paths:\\
\begin{itemize}
\item[(i)] the lower path: measure the particle with the randomness generating measurement, $\mathcal{M}_{A}$. Denote the average post-measurement state as $\rho'$; \\
\item[(ii)] the upper path: no operation is performed.
\end{itemize}
\item[(3)] Perform a subsequent testing measurement, $\mathcal{M}_{B}$, on particles evolved after step 2 and record the measurement outcomes for each path, respectively.
	
\item[(4)] Repeat steps $1-3$ for sufficiently many times. The measurement-outcome distributions of $\mathcal{M}_{B}$ corresponding to $\rho$ and $\rho'$ are denoted as $\mathbf{q}=\{q(j|B; \rho)\}$,  and $\mathbf{q}'=\{q'(j|B; \rho')\}$, respectively.

\item[(5)] Estimate the disturbance of $\mathcal{M}_{A}$ to $\rho$ (and  $\mathcal{M}_{B}$) from the distance between the distributions $\mathbf{q}$ and $\mathbf{q}'$ and estimate the amount of generated randomness.
\end{itemize}

In step.2,  every  particle prepared by  source  is randomly sent to the upper or the lower paths  according to a pre-fixed probability distribution, which is controlled with a true random number.  Let the total number $n$ of prepared particles  be sufficiently large,  and $n_{u}$ be the number of particles sent  to upper path for the estimation of $\mathbf{q}$.
To randomly choose $n_{u}$ from $n$,  one need $t(s)=\lceil \log_{2} \frac{n!}{n_{u}! (n-n_{u})!}\rceil$ bits of quantum randomness. Typically, $n_{u}\ll n$, for instance,   $n_{u}$ can be chosen as $n_{u} = \lceil\sqrt{n}\rceil$, thus one has $t(s)\rightarrow \sqrt{n}\log_{2} \sqrt{n}$. The averaged true randomness consumed in each run is thus $\frac{t(s)}{n-n_{u}}\approx \frac{\log_{2}\sqrt{n}}{\sqrt{n}}$, which  tends to zero when $n\rightarrow \infty$. Thus,  a  negligible fraction  of the
generated randomness is sufficient to feed back as  new seeds.

\subsection{Security Assumptions}
In the protocol, we do not trust the state source \emph{a priori}. We simply take the following assumptions on the measurements:
\begin{enumerate}
  \item \emph{Independence between devices}:  the randomness-generating measurement and the testing measurement are mutually independent.
 \item  \emph{L\"{u}ders' rule}: the randomness-generating measurement obeys  L\"{u}ders' rule.
\end{enumerate}
The independence assumption requires that measurements $\mathcal{M}_A$ and  $\mathcal{M}_B$ are  trusted to be  non-malicious.  That is,   measurement $\mathcal{M}_B$ does not  apply selective measurements by using the information of whether $\mathcal{M}_{A}$ is performed or not.  
  The second assumption is about the realization of measurement $\mathcal{M}_{A}$, namely, it is a L\"uders' instrument~\cite{1950A, 936140} with state updating $\rho\to\rho'=\sum_{i}\sqrt{M_{i}}\rho \sqrt{M_{i}}$.
  Generally, the post-measurement state $\rho'$ (and thus the disturbed data $\mathbf{q}'$) depends on the realization of $\mathcal{M}_{A}$ and is given as $\sum_{i}{\rm U}_{i}\sqrt{M_{i}}\rho \sqrt{M_{i}}{\rm U^{\dagger}}_{i}$, where the  unitary operations $\{{\rm U}_{i}\}$ are  determined by the realization. As a realization of great interest,  L\"{u}ders' instrument does not break quantum coherence more than the necessary and disturbs  unknown input $\rho$ to the minimal extent~\cite{936140,PhysRevA.63.062305, PhysRevLett.86.1366, Winter2014Coding, 2007Coding}. As will be shown in what follows,  this minimal state change is seen as  state collapse that leads to the generation of quantum randomness. Besides these two main assumptions, we will also show that our protocol  can be further optimized with more  assumptions on the knowledge of the measurements.


\section{Estimating Randomness under  Classical Adversary}
 To establish the connection between quantum randomness and disturbance effect, our essential tools are kinds of  uncertainty-disturbance relations~\cite{sun2022disturbance}, which indicate that the uncertainty in the measurement of $\mathcal{M}_{A}$ upper-bounds its disturbance effect in the quantum state and  a following quantum measurement and has been used to estimate quantum coherence~\cite{PhysRevA.106.042428}.

 In this section, we deal with  $\operatorname{H}^{C}_{min}$, and the uncertainty of measurement $\mathcal{M}_{A}$ is defined as
\begin{eqnarray}\label{defun}
\delta_{\mathcal{A}; \rho}:=\sqrt{1-\textstyle\sum_{i}p^{2}(i|A; \rho)}, \nonumber
\end{eqnarray}
where $\mathbf{p}=\{p(i|A; \rho)\}$ specifies the outcome distribution of measurement  $\mathcal{M}_{A}$. This measure is related to the collision entropy, $\operatorname{H}_{2}(A)=-\log(1-\textstyle\delta^{2}_{\mathcal{A};\rho})$, where $\operatorname{H}_{2}(A):=-\log \textstyle\sum_{i}p^{2}(i|A; \rho)$. After measuring $\mathcal{M}_{A}$, the initial state ensemble $\rho$ is transferred into $\rho'$. The degree of state collapse is quantified with trace distance,
\begin{eqnarray}
\operatorname{D}_{\rho,\rho'}=\textstyle\frac{1}{2}\operatorname{tr}|\rho-\rho'|,   \nonumber
\end{eqnarray}
which can be estimated by total-variance (TV) distance between distributions arising from  performing $\mathcal{M}_{B}$ on $\rho$ and $\rho'$, respectively. The TV distance is
\begin{eqnarray}
\operatorname{D}_{A\to B;\rho}=\textstyle\frac{1}{2}
\textstyle\sum_{j}|q(j|B; \rho)-q'(j|B; \rho')|,\nonumber
\end{eqnarray}
which directly quantifies the disturbance introduced by $\mathcal{M}_{A}$ to $\mathcal{M}_B$. According to the data processing inequality, we have
\begin{eqnarray}
\textstyle \operatorname{D}_{\rho,\rho'}\geq \textstyle \operatorname{D}_{A\to B;\rho}.  \nonumber
\end{eqnarray}
Note that $\operatorname{D}_{A\to B;\rho}=0$ happens only when  $\rho=\rho'$  or $\rho-\rho'$  is perpendicular to all the elements of $\mathcal{M}_{B}$ simultaneously when $\rho\neq \rho'$, such state lie in a space of  measure zero. This implies that the protocol can use almost all the states  to verify the state collapse and hance to generate quantum randomness.

\subsection{QRNG Using POVMs}
At first, we consider the  case with the assumptions of \emph{Independence between devices} and \emph{L\"{u}der's rule} only and take  $\mathcal{M}_{A}$ and $\mathcal{M}_{B}$ as  POVMs whose elements are unknown.

\begin{Lemma}[Uncertainty-disturbance  lemma for POVMs]
Given a general measurement, $\mathcal{M}_{A}$, which follows L\"{u}ders' rule, suppose it is applied to a state, $\rho$, and the average post-measurement state is $\rho'$. The uncertainty in measurement outcomes is lower-bounded by its disturbance effect in the measured state to a following  measurement, $\mathcal{M}_{B}$,
  \begin{eqnarray}
\delta_{A; \rho}\geq \operatorname{D}_{\rho,\rho'} \geq  \operatorname{D}_{A\to B; \rho}, \label{ud}
\end{eqnarray}
\end{Lemma}
We leave the proof of this result in Supplementary Information(SI). Based on this lemma, we come to our first main result.

\begin{Theorem}[Quantum randomness based on the disturbance effect of  POVMs]\label{Lemma:UncertainDist}
For a measurement, $\mathcal{M}_{A}$, which follows L\"{u}ders' rule, and a subsequent measurement, $\mathcal{M}_{B}$, the disturbance that $\mathcal{M}_{A}$ causes in a state to $\mathcal{M}_{B}$  implies a lower bound on the randomness generated by measuring $\rho$ with $\mathcal{M}_{A}$,
 \begin{eqnarray}
\operatorname{H}^{C}_{min}(A|\rho)\geq -\log \left(\textstyle\frac{1}{2}+\textstyle\frac{1}{2}\sqrt{1- 2\operatorname{D}^{2}_{A\to B, \rho}}\right). \label{rand1}
\end{eqnarray}
\end{Theorem}

Here, we give a sketch of the proof. We leave the details
in SM.
\begin{proof}
We first consider a pure state, $\rho=|\phi \rangle\langle \phi|$. By using Eq.~\eqref{ud}, Eve's guessing probability is
\begin{equation}
\operatorname{G}_{A|\phi}=\max\{p(0|A; \phi), p(1|A; \phi)\}. \nonumber
\end{equation}
Using Lemma~\ref{Lemma:UncertainDist}, we have
\begin{equation}
\sqrt{1-p^{2}(0|A; \phi)- [1-p(0|A; \phi)]^{2}}\geq \operatorname{D}_{ \phi, \phi'}\geq \operatorname{D}_{A\to B; \phi}. \nonumber
\end{equation}
The disturbance,  $\operatorname{D}_{A\to B; \phi} $ (and $\operatorname{D}_{ \phi, \phi'}$) implies upper bound for $\operatorname{G}_{A|\phi}$,
\begin{equation}
\textstyle\frac{1}{2}\leq \operatorname{G}_{A|\phi}\leq \textstyle\frac{1}{2}+\textstyle\frac{1}{2}\sqrt{1-2\operatorname{D}^{2}_{A\to B; \phi}}. \nonumber
\end{equation}
For a general mixed state case, we first decompose $\rho$ with a convex combination of pure states. By applying the result to every element  pure state in the decomposition  and using a convexity argument, we arrive at Eq.~\eqref{rand1}.
\end{proof}

\subsection{QRNG Using  Projection  Measurement}
As the second illustration, we assume  the measurement $\mathcal{M}_{A}$  to be   projective additionally but not necessarily characterized, namely, the elements of measurement are unknown. This  assumption  allows us to employ a tighter uncertainty-disturbance relation  introduced in Refs.~\cite{ucb, sun2022disturbance}.

\begin{Lemma}[Uncertainty-disturbance relation for unknown projection measurement]
In a sequential measurement scheme, up to a factor of  $\frac{\sqrt{2}}{2}$, the uncertainty of measuring a state, $\rho$, with a projection  measurement, $\mathcal{M}_{A}$, would be no less than its disturbance effect in the measured state to a following  measurement, $\mathcal{M}_{B}$, \begin{eqnarray}
\textstyle \frac{\sqrt{2}}{2} \delta_{A; \rho}\geq \operatorname{D}_{\rho;\rho'}  \geq \operatorname{D}_{A\to B;\rho}.
\label{undis2}
\end{eqnarray}
\end{Lemma}

Immediately, we have the following randomness estimation result.

\begin{Theorem}
Quantum randomness generated by a projection measurement is lower-bounded by its disturbance effect to a following measurement, namely,
\begin{eqnarray}
\operatorname{H}^{C}_{min}(A|\rho)\geq -\log \left(\textstyle\frac{1}{2}+\textstyle\frac{1}{2}\sqrt{1- 4\textstyle \operatorname{D}^{2}_{A\to B, \rho}}\right).
\label{rand2}
\end{eqnarray}
\end{Theorem}
The proof follows essentially the same route as in proving Theorem~\ref{Lemma:UncertainDist}.
\subsection{QRNG Using  von Neumann Measurements}
In the last, we consider another case in which measurement $\mathcal{M}_{A}$ and $\mathcal{M}_{B}$ are completely trusted to be rank-one projection measurements, or von Neumann
measurements,  as $\mathcal{M}_{A}=\{|i\rangle\langle i|\}$ and $\mathcal{M}_{B}=\{|b_{j}\rangle\langle b_{j}|\}$. There exists an optimal uncertainty-disturbance relation~\cite{sun2019disturbance} for two-dimensional systems.  Denote  the overlaps between  eigenvectors of the measurements as $c_{ij}=|\langle i |b_{j}\rangle|^{2}$, we have

\begin{Lemma}[Uncertainty-disturbance  relation for trusted von Neumann measurements]
In a sequential measurement scheme $\mathcal{M}_{A}\to \mathcal{M}_{B}$ where both measurements are trusted von Neumann measurements, up to a factor $\frac{1}{2}\delta_{\mathcal{A}:\mathcal{B}}:=\sum_{j}\sqrt {1-\sum_{i}c^{2}_{ij}}$, the uncertainty in the measurement results of $\mathcal{M}_{A}$ would be no less than its disturbance effect to $\mathcal{M}_{B}$,
\begin{eqnarray}\label{g2}
\textstyle\frac{1}{2}\delta_{\mathcal{A}:\mathcal{B}}\delta_{\mathcal{A}; \rho}\geq \operatorname{D}_{A\to B;\rho}.
\end{eqnarray}
\label{Lemma:RelationTrusted}
\end{Lemma}

We can also apply the uncertain-disturbance relation to derive a randomness estimation from the disturbance effect. For the binary case, $\textstyle\delta_{\mathcal{A}:\mathcal{B}}=2\sqrt{2c_{00}(1-c_{00})}\leq \sqrt{2},$  implying that Eq.~\eqref{g2} gives a tighter estimation than Eq.~\eqref{undis2}. We have the following stronger randomness estimation result.

\begin{Theorem}
Perform trusted von Neumann measurements sequentially, $\mathcal{M}_{A}\to \mathcal{M}_{B}$, the disturbance introduced by $\mathcal{M}_{A}$ to $\mathcal{M}_{B}$ implies a lower-bound on the randomness generated by $\mathcal{M}_{A}$,
\begin{eqnarray}
\operatorname{H}_{min}(A|\rho)\geq -\log \left(\textstyle\frac{1}{2}+\textstyle\frac{1}{2}\sqrt{1- 4
\textstyle \operatorname{\tau}^{2}_{A\to B, \rho}}\right),
\label{rand3}
\end{eqnarray}
where we call $\tau_{A\to B, \rho}:=\frac{\sqrt{2}\operatorname{D}_{A\to B, \rho}}{\delta_{\mathcal{A}:\mathcal{B}}}$ the modified disturbance.
\end{Theorem}

\begin{figure}
\begin{center}
 \includegraphics[width=0.6\textwidth]{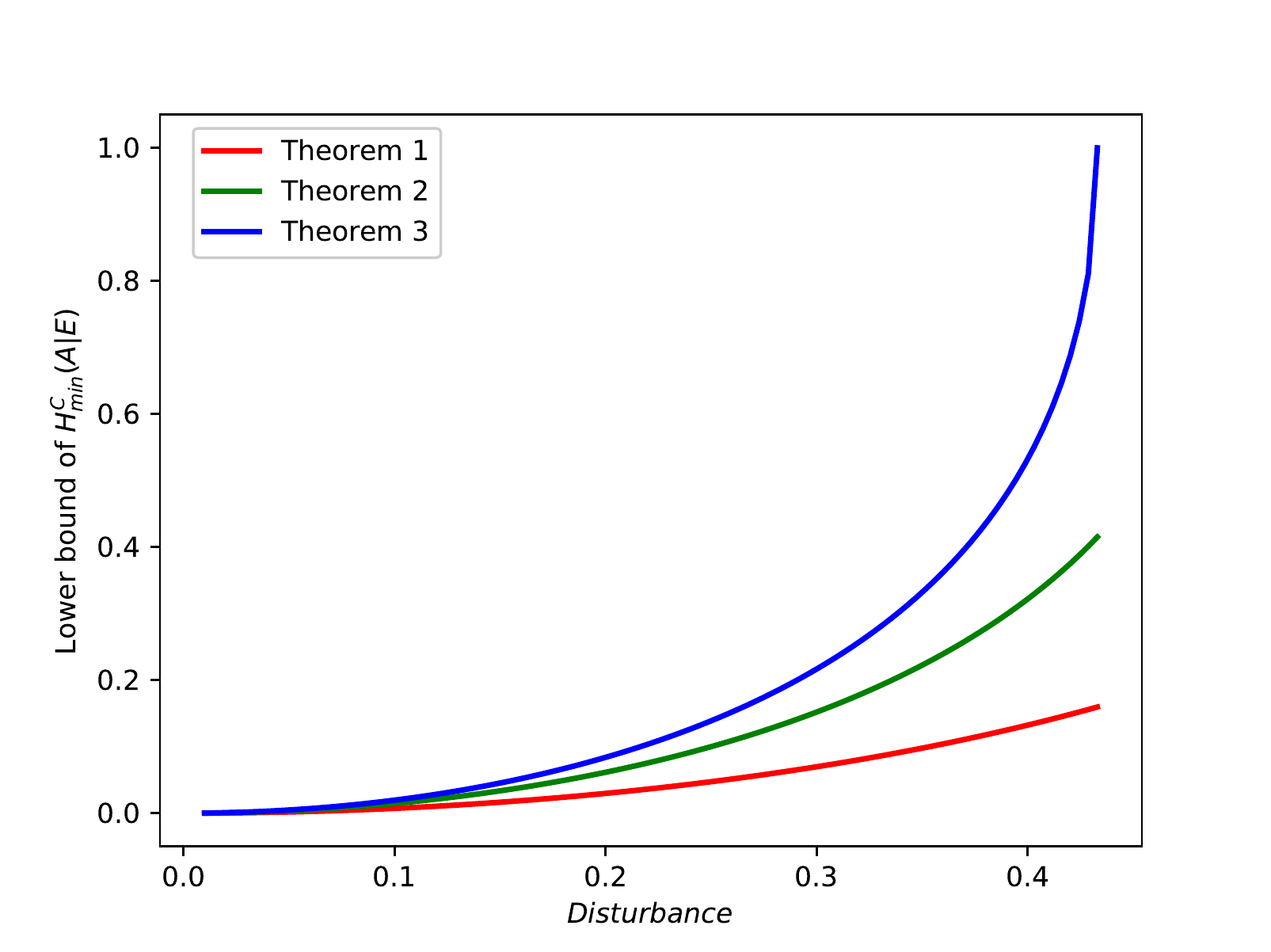}
\end{center}
\vspace{+0cm}
\caption{ Three lower bounds  under different assumptions  are drawn as functions of disturbance when $|\langle 0|b_{0}\rangle|^{2}=0.75$.    }\label{fig2}
\end{figure}

To illustrate the advantage of the modification with $\delta_{\mathcal{A}:\mathcal{B}}$, consider a condition where the measurement $\mathcal{M}_{A}$ is very close to $\mathcal{M}_{B}$. Then, $\operatorname{D}_{A\to B; \rho}$ is close to zero and so are the lower bounds of Eq.~\eqref{rand1} and \eqref{rand2}, even if $\operatorname{D}_{\rho; \rho'}$ is significantly large. The triviality is overcome by  $\tau_{A\to B, \rho}$ as  the denominator    $\delta_{\mathcal{A}:\mathcal{B}}$ also is close to zero, which  amplifies the disturbance such that a large amount of quantum randomness can be verified.  We compare the three lower bounds in Fig.~\ref{fig2}, which clearly shows that the information about measurements can significantly optimize the  protocol's performance. We note that,  when the basis of $\mathcal{M}_{B}$ is taken as the eigenvectors of $\rho-\rho'$, the maximum disturbance is acquired as    the degree of state collapse, $i.e.$, $\max_{B}\operatorname{D}_{A\to B, \rho}=\max_{B}\tau_{A\to B, \rho}=\operatorname{D}_{\rho,\rho'}$, and the maximum experimentally  accessible  randomness generation is given as functions of  state collapse.  We thus call our method a direct way based on state collapse.  This is different from other protocols where the randomness are bounded in terms of functions of some non-classical quantities, such as nonlocality and contextuality.

Let us return to  the example mentioned in the very beginning,  and let  $\mathcal{M}_{A}=\sigma_{z}$ and $\mathcal{M}_{B}=\sigma_{x}$.   When $\rho$ is $\frac{\sqrt{2}}{2}(|0\rangle+|1\rangle)$,   then   $\operatorname{D}_{A\to B; \rho}=\frac{1}{2}$, $\tau_{A\to B; \rho}=\frac{1}{2}$ (with $\delta_{\mathcal{A}:\mathcal{B}}=\sqrt{2}$). Then, both  Eq.(\ref{rand2}) and Eq.~\eqref{rand3} provide  optimal  lower bound as $1$, and $1-\log (1+\sqrt{\frac{1}{2}})$ by Eq.(\ref{rand1}).  When $\rho=\frac{1}{2}|0\rangle\langle0|+\frac{1}{2}|1\rangle\langle1|$,  disturbance is zero  and so are the lower-bounds of all  the verifications.

To conclude this section,  we  analyse  the tolerance of our protocols to  the noise in real   measurements  that do not assume L\"{u}ders' instrument or projective measurement perfectly. In Theorem.1 and Theorem.2,  nothing is assumed  on the characterization on $\mathcal{M}_{B}$ except its independence on $\mathcal{M}_{A}$.  To analyse the noise from   $\mathcal{M}_{A}$,  we  denote  the channel corresponding to  the real measurement  approximating $\mathcal{M}_{A}$  by $\Lambda_{re}(\cdot)$, and the channel corresponding to the ideal measurement $\mathcal{M}_{A}$ assuming   L\"{u}ders' instrument or  projection measurement  by  $\Lambda_{id}(\cdot)$~\cite{PhysRevA.71.062310}. They  update an input state $\rho$ respectively into  $\Lambda_{re}(\rho)$ and  $\Lambda_{id}(\rho)=\rho'$.
  The total state change,   quantified by the distance ${\rm D}_{\rho, \Lambda_{re}(\rho)}$ between $\rho$ and the post-measurement state $\Lambda_{re}(\rho)$ in experiment,  can be partitioned into parts: namely,  the   one  used to generate randomness  ${\rm D}_{\rho, \rho'}$ and the one due to the imperfection of $\mathcal{M}_{A}$, namely,   ${\rm D}_{\rho', \Lambda_{re}(\rho)}$. By the triangle-inequality of trace-distance  we have
 \begin{eqnarray}
{\rm D}_{\rho, \rho'}\geq {\rm D}_{\rho, \Lambda_{re}(\rho)}-{\rm D}_{ \Lambda_{re}(\rho), \rho'}. \label{mod1}
\end{eqnarray}
We  can  assume the maximum noise over any input state  $\varepsilon_{A}:=\max_{\rho}{\rm D}_{\Lambda_{re}(\rho), \rho'}$ to be  a known small quantity~\cite{PhysRevLett.124.080401} for a trusted L\"{u}ders'  instrument of $\mathcal{M}_A$  as,    by the state of art platforms,  the fidelity between the two channels $\Lambda_{re}$ and $\Lambda_{id}$ can reach  a fidelity as high as $96\%$~\cite{PhysRevApplied.6.014014}. Performing measurement $\mathcal{M}_{B}$ on $\rho$ and    $\Lambda_{re}(\rho)$ respectively,  we obtain a disturbance ${\rm D}^{(re)}_{ {\rm A\to B}, \rho}$, which is defined by the distance between the resulting  statistics, that lower-bounds the state distance in experiment, i.e., ${\rm D}_{\rho, \Lambda_{re}(\rho)} \geq {\rm D}^{(re)}_{ A\to B, \rho}.$
Together with  Eq.(\ref{mod1}),   we  have  a lower bound on the state collapse $ {\rm D}_{\rho, \rho'}$ as
 \begin{eqnarray}
{\rm D}_{\rho, \rho'}\geq  {\rm D}_{\rho, \Lambda_{re}(\rho)}-{\rm D}_{\Lambda_{re}(\rho), \rho'}\geq {\rm D}^{(re)}_{{\rm A}\to {\rm B}, \rho}-\varepsilon_{A}\equiv{\rm D}^{\varepsilon}_{{\rm A}\to {\rm B}, \rho}.
\end{eqnarray}
  Replacing  ${\rm D}_{\rm A\to B, \rho}$ in Theorem.1 and Theorem.2  with  ${\rm D}^{\varepsilon_{A}}_{\rm A\to B, \rho}$, we extend these theorems to a practical case. For example, the Theorem.1 is given as
$$\operatorname{H}^{C}_{min}(A|\rho)\geq -\log \left(\textstyle\frac{1}{2}+\textstyle\frac{1}{2}\sqrt{1- 2({\rm D}^{\varepsilon_{A}}_{\rm A\to B, \rho})^{2}}\right).$$

In the Theorem.3, we additionally  assume that both $\mathcal{M}_{A}$ and $\mathcal{M}_B$ are rank-one projection measurements.  The imperfection of $\mathcal{M}_{A}$ can be incorporated in $\varepsilon_{A}$. For  $\mathcal{M}_B$,  we define the noise of measurement $\mathcal{M}_B$ as $\varepsilon_{B}:=\frac{1}{2}\max_{\rho} \sum_{i} |{\rm Tr}(\rho\cdot |b_{i}\rangle \langle b_{i}|)- {\rm Tr}(\rho\cdot M^{(re)}_{b_i})|$ with $M^{(re)}_{b_i}$ being  the measurement element  approximating the ideal one $|b_{i}\rangle \langle b_{i}|$.  This noise generally can be seen as a small quantity that is no more  than $5\%$ for  qubit system~\cite{2021A}. After incorporating two errors $\varepsilon_{A}$ and $\varepsilon_{B}$ (See SI),  we have
$\frac{1}{2}\delta_{\mathcal{A}:\mathcal{B}}\delta_{\mathcal{A}; \rho}\geq {\rm D}^{\varepsilon_{A}, \varepsilon_{B}}_{\rm A\to B, \rho}$ with
 ${\rm D}^{\varepsilon_{A}, \varepsilon_{B}}_{\rm A\to B, \rho}:={\rm D}^{(re)}_{\rm A\to B, \rho}-\varepsilon_{A}- 2\varepsilon_{B}$. The effects of the noise of practical measurements on the rate of generated randomness can be accounted  for by  replacing $\operatorname{D}_{A\to B, \rho}$ in Theorem.3 with ${\rm D}^{\varepsilon_{A}, \varepsilon_{B}}_{\rm A\to B, \rho}$.

\section{Quantum Randomness in  the asymptotic Limit}
In this section, we use another  uncertainty-disturbance relation to estimate quantum randomness in the asymptotic limit of an infinite data size and compare our protocol with the one based on uncertainty relation.

The uncertainty of measurement $\mathcal{M}_{A}$ is defined as the Shannon entropy $\rm H(\bf{p})$, the disturbance in state is defined as quantum relative entropy   $S(\rho\|\Delta(\rho))$, and  the disturbance effect to a subsequent measurement is defined as the classical relative entropy, or, Kullback-Leibler divergence, $\operatorname{H}(\mathbf{q}\|\mathbf{q}'):=-\sum_{i}q(j|B; \rho)\log\frac{q(j|B; \rho)}{q'(j|B; \rho')}$.
\begin{Lemma}[Uncertainty-disturbance  lemma for  projection measurement]
For one projection measurement $\mathcal{M}_{A}$, its uncertainty quantified by $\rm H(\bf{p})$ is  no less than its disturbance effect  in the measured  state and a following measurement~\cite{sun2022disturbance}:
\begin{eqnarray}
\rm H(\mathbf{p})\geq  S(\rho\|\Delta(\rho)) \geq \rm{H}(\bf{q}\|\bf{q}').
\end{eqnarray}
\end{Lemma}
With the definition of quantum randomness given by Eq.~\eqref{cr}, we immediately have a bound on  $\operatorname{H}^{Q,C}_{min, asy}(A|E)$~\cite{sun2022disturbance} with the operational meaning of the disturbance effect.

\begin{Theorem}
In the asymptotic limit, quantum randomness generated from a measurement, $\mathcal{M}_A$, can be lower-bounded by its disturbance effect to a subsequent measurement,
\begin{eqnarray}
\operatorname{H}^{Q, C}_{min, asy}(A|E)\geq \operatorname{H}(\mathbf{q}\|\mathbf{q}'), \label{ud3}
\end{eqnarray}
\end{Theorem}
The lower bound of $\operatorname{H}^{Q}_{min, asy}(A|E)$ is obvious. For $\operatorname{H}^{C}_{min, asy}(A|E)$, we have used that, for any decomposition of $\rho=\{r_{n}, \phi_{n}\}$, one has  $\textstyle\sum_{n} r_{n}\cdot \operatorname{H}(\mathbf {p_{\phi_n}})\geq \sum_{n} r_{n}\cdot \operatorname{H}(\mathbf{q}_{\phi_{n}}\|\mathbf{q}'_{\phi'_{n}})\geq \operatorname{H}(\mathbf{q}\|\mathbf{q}')$.

With Lemma~\ref{Lemma:RelationTrusted}, we can provide another lower bound on $\operatorname{H}^{C}_{min,asy}(A|E)$.

\begin{Theorem}
The quantum randomness generated by a projection measurement, $\mathcal{M}_{A}$, is lower-bounded by its modified disturbance in a subsequent projection measurement, $\mathcal{M}_{B}$,
\begin{eqnarray}
\operatorname{H}^{C}_{min,asy}(A|E)\geq  4\tau^{2}_{A\to B, \rho}.
\label{rand4}
\end{eqnarray}
\end{Theorem}
This bound  is always better than the above one in the sense that for any pair of $\mathbf{q}$ and $\mathbf{q}'$,  $ 4\tau^{2}_{A\to B, \rho}\geq \operatorname{H}(\mathbf{q}\|\mathbf{q}')$.

\begin{figure}
\begin{center}
\includegraphics[width=0.6\textwidth]{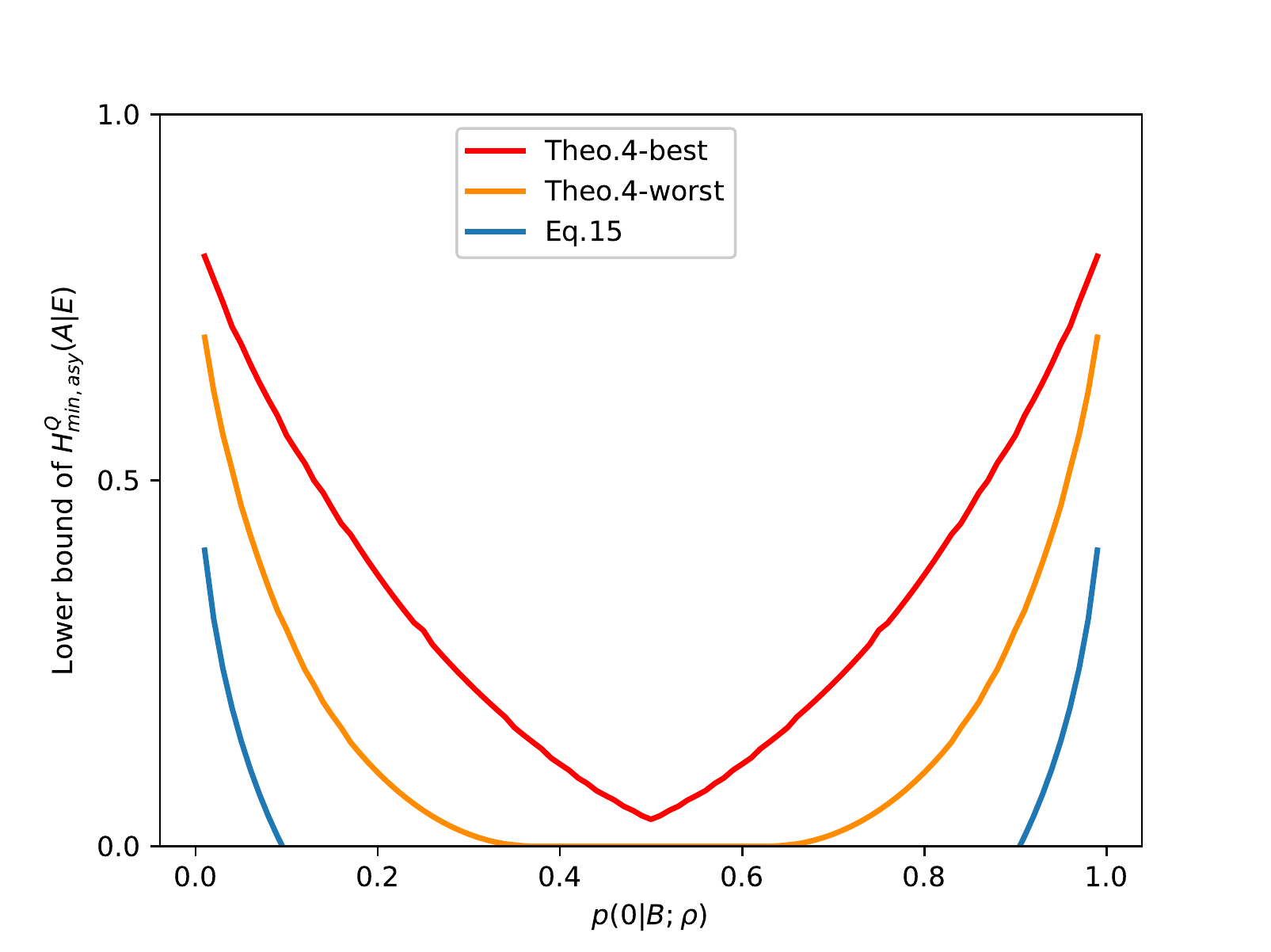}
\end{center}
\vspace{+0cm}
\caption{ Comparison between the Lower-bounds implied by  uncertainty relation and  uncertainty-disturbance relation,  respectively. Set $c=0.62$ and three lines  are drawn. The lower bound in Eq.(\ref{mu}) depends only on $\mathbf{q}$.  The lower bound from Theorem.4 depends also on $\mathbf{p}$. We minimize and maximize the bound over $\mathbf{p}$ in qubit space, respectively. It can be seen that even the worst case is  better than the one based on uncertainty relation.      }\label{fig3}
\end{figure}

\section{In comparison with the Verification based on Uncertainty Relation}
Uncertainty and  disturbance effect are two  fundamental  traits of quantum measurement.  Usually, their presence is separately found in the state preparation uncertainty relation and the measurement uncertainty relation. Here, we link the two quantities by the uncertainty-disturbance relations.

Preparation uncertainty states that a quantum system cannot be  prepared in the eigenstates of two incompatible observables simultaneously. In the presence of side information as in our discussions, the uncertainty relation can be harnessed to verify generation of quantum randomness~\cite{PhysRevA.90.052327},
\begin{eqnarray}
\operatorname{ H}^{Q}_{min, asy}(A|E)\geq -\log c -\operatorname{H}_{\frac{1}{2}}(\operatorname{B}). \label{mu}
\end{eqnarray}
where $c=\max_{i,j}c_{ij}$ is the overlap between the two observables and $\operatorname{H}_{\frac{1}{2}}(\operatorname{B})$ is $\textstyle\frac{1}{2}-$R\'{e}nyi  entropy of $\mathbf{q}$. Clearly, the measurement  incompatibility, $i.e.$, $\{c_{ij}\}$ is the key ingredient.

Following a different route, our protocol uses the disturbance effect that manifests itself in sequential measurements. It requires L\"{u}ders' rule instead of the measurement incompatibility in terms of the state overlap. When comparing the lower bounds, as $\textstyle q'(j|B; \rho)=\operatorname{ tr}(\rho'|b_{j}\rangle\langle b_{j}|)$$=\sum_{i} c_{ij}p(i|A; \rho)$ $\leq c$, Eq.~\eqref{ud3} always provides a better estimation than Eq.~\eqref{mu}, as
\begin{eqnarray}
\operatorname{H}(\mathbf{q}\|\mathbf{q}')&=&\textstyle\sum_{i}q(j|B;\rho)\log\frac{q(j|B; \rho)}{q'(j|B; \rho')}\nonumber \\&\geq&\textstyle-\operatorname{H}(\operatorname{B})-\log c \nonumber \\
&\geq& -\operatorname{H}_{\frac{1}{2}}(\operatorname{B})-\log c,
\end{eqnarray}
where we use $\operatorname{H}(\operatorname{B})\leq \operatorname{H}_{\frac{1}{2}}(\operatorname{B})$.  Different from the bound in Eq.~\eqref{mu}, the bound in Eq.~\eqref{ud3} depends additionally on $\mathbf{p}$.
For a graphical comparison, we restrict ourselves to the case where both $\mathcal{M}_{A}$ and $\mathcal{M}_{B}$ are trusted projection measurements.  We   maximize and minimize $\operatorname{H}(\mathbf{q}\|\mathbf{q}')$ from  Eq.~\eqref{ud3} over $\mathbf{p}$ in the qubit space, which corresponds to the best and the worst randomness estimation result. We have shown the figures in Fig.~\ref{fig3}.  Even the worst-case lower bound  is significantly better than the estimation result obtained from the uncertainty relation.

\section{Discussion and Conclusion}
Fundamentally, quantum randomness always comes from  the unpredictability of state collapse induced by measurement. This  unpredictability  has previously   been  ensured with quantum features such as nonlocality, non-contextuality, state distinguishability, $et$ $al.$~\cite{Ekert1991Quantum, PhysRevLett.95.010503, colbeck2011quantum, PhysRevLett.108.100402, 2009Masanes, 2010Random, 2018Device,  PhysRevA.86.062109, 2018Corrigendum, PhysRevLett.119.240501, PhysRevApplied.7.054018,  PhysRevA.84.034301, PhysRevLett.114.150501,  2016guo}, and the demonstration of the properties often require additional assumptions on devices.  In this paper, we have   bridged in a direct way   the two fundamental concepts, $i.e.$,  randomness generation and the degree of state collapse, with a proposed  a QRNG, where the state collapse is  estimated with  disturbance in measurement.  The protocol enables us to estimate   quantum randomness in the various scenarios, $i.e.$,   under classical adversary,   in asymptotic limit under the classical and the quantum adversaries,  respectively.  The protocol employs trivial mathematics. Quantum  randomness is verified  once the disturbance is measured, and one does not need to  take an optimization over the unknown parameters of devices. Our protocol also shows high efficiency and  outperforms the protocol based on uncertainty relation.   These merits  are greatly beneficial for the real-time QRNGs.    Therefore, we provide new insights on  the estimation of  quantum randomness and the practical design of QRNGs.

The uncertainty-disturbance relation Eq.(\ref{ud}),  though not claimed as the main result in this paper, actually may have an independent interest in the field of channel coding. It can be seen as  a generalized gentle measurement lemma~\cite{Winter2014Coding}  to the case where the measurement  is not necessarily \emph{gentle} as it is not trivial even the uncertainty is significantly large, namely, the measurement is not gentle.  It may find applications in the field where the lemma has played a key role.

\textbf{References}

\bibliographystyle{naturemag}
\bibliography{citation}

\newpage


\section*{Acknowledgments}
 L.L.Sun and S.Yu would like to thank Key-Area Research and Development Program of Guangdong Province Grant No. 2020B0303010001.  J.F. and Z.D.L is supported by Guangdong Provincial Key Laboratory Grant No.2019B121203002 and Key-Area Research and Development Program of Guangdong Province Grant No. 2020B0303010001 and No. 2019ZT08X324.
\section*{Supplementary Information}

\subsection{Proof of Lemma 1}
The measurements $\mathcal{M}_A$ and $\mathcal{M}_B$   are general positive operator-valued measures (POVM).  The set of POVMs for   $\mathcal{M}_A$ is represented as $\{\operatorname{M}_{i}\}$, $\sum_{i}\operatorname{M}_{i}=\operatorname{I}$.  With  assumption in the main text, the post measurement state   after measuring $\mathcal{M}_A$ on $\rho=\sum_{n} r_{n}\cdot |\phi_{n}\rangle\langle \phi_{n}|$ is $\rho'=\sum_{i}\sqrt{\operatorname{M}_{i}}\cdot \rho \cdot \sqrt{\operatorname{M}_{i}}$.
We have
   \begin{eqnarray*}
\operatorname{D}_{\rho,\rho'}=\textstyle\frac{1}{2}\operatorname{tr}|\rho-\rho'|& = &\textstyle\frac{1}{2}\operatorname{tr}|\textstyle\sum_{n}r_{n}|\phi_{n}\rangle\langle \phi_{n}|- \textstyle\sum_{n, i}r_{n}\textstyle\sqrt{\operatorname{M}_{i}}\cdot|\phi_{n}\rangle\langle \phi_{n}|\cdot\textstyle\sqrt{\operatorname{M}_{i}}|\nonumber \\
&\leq &\textstyle\sum_{n}r_{n}\operatorname{tr}\big ||\phi_{n}\rangle\langle\phi_{n}|- \textstyle\sum_{i}\sqrt{ M_{i}}\cdot|\phi_{n}\rangle\langle \phi_{n}|\cdot\textstyle\sqrt{ \operatorname{M}_{i}}\big | \nonumber\\
&\leq &\textstyle\sum_{n}r_{n}\sqrt{1-\textstyle\sum_{i}\langle\phi_{n}|\textstyle\sqrt{\operatorname{M}_{i}}|\phi_{n}\rangle\langle \phi_{n}|\textstyle\sqrt{ \operatorname{M}_{i}}|\phi_{n}\rangle}\\
&\leq& \textstyle\sqrt{1-\textstyle\sum_{n,i}r_{n}\langle\phi_{n}|\textstyle\sqrt{\operatorname{M}_{i}}|\phi_{n}\rangle^{2}}\\
&=& \sqrt{1- \textstyle\sum_{k}r_{k}\textstyle\sum_{n,i}(\textstyle\sqrt{r_{n}}\langle\phi_{n}|\sqrt{\operatorname{M}_{i}}|\phi_{n}\rangle)^{2}}\\
&\leq&\textstyle\sqrt{1-(\textstyle\sum_{n,i}\sqrt{r_{n}}\sqrt{r_{n}}\langle\phi_{n}|\sqrt{\operatorname{M}_{i}}|\phi_{n}\rangle)^{2}}\\
&\leq&\textstyle\sqrt{1-\textstyle\sum_{i}[\operatorname{tr}(\sum_{n}r_{n}|\phi_{n}\rangle\langle\phi_{n}|\cdot\sqrt{\operatorname{M}_{i}})]^{2}}\\
&\leq&\textstyle\sqrt{1-\textstyle\sum_{i}[\operatorname{tr}(\rho \operatorname{M}_{i})]^{2}}=\delta_{A; \rho}
\end{eqnarray*}
where trace-norm $\operatorname{tr}|A|=\operatorname{tr}\sqrt{AA^{\dagger}}$ and  the second inequality is due to triangle inequality, the third inequality is due to that  $\operatorname{tr}|\rho-\sigma|\leq 2\sqrt{1-\operatorname{tr}(\rho\cdot \sigma) }$ holds for two arbitrary quantum states $\rho$ and $\sigma$ and the fifth inequality is due to $(\sum_{i}a^{2}_{i})(\sum_{j}b^{2}_{j})\geq (\sum_{i}a_{i}b_{i})^{2}$ and the sixth inequality is due to $\operatorname{M}_{i}\leq \sqrt{\operatorname{M}_{i}}$.

By $\{p(j|B; \rho)\}$, $\{p(j|B; \rho')\} $ we specify the outcome  distributions  from a general measurement  $\mathcal{M}_B$ performed on $\rho$ and $\rho'$ respectively. We  then have
  \begin{eqnarray*}
\textstyle\sqrt{1-\textstyle\sum_{i}p^{2}_{i}}\geq \frac{1}{2}\operatorname{tr}|\rho-\rho'|\geq \frac{1}{2} \sum_{i} |q(j|B; \rho)-q(j|B; \rho')|
\end{eqnarray*}

\subsection{Proof of Theorem~1}
As a warmup,   we first  relate the correctly guessing probability to disturbance effect for the case of  performing $\mathcal{M}_{A}$ on a pure state, namely, $\rho=|\phi \rangle\langle \phi|$.
    Immediately,
  $$\operatorname{G}_{A|\phi}=\max\{ p(0|A; \phi), 1- p(0|A; \phi)\}.$$
It follows from  Eq.(\ref{ud}) that:
  \begin{eqnarray*}
\sqrt{1-p^{2}(0|A; \phi)- [1-p(0|A; \phi)]^{2}}\geq \operatorname{D}_{ \phi, \phi'},
\end{eqnarray*}
Unifying them yields bounds for $\operatorname{G}_{A|\phi}$ as
 \begin{eqnarray}
\textstyle\frac{1}{2}\leq \operatorname{G}_{A|\phi}\leq \textstyle\frac{1}{2}+\textstyle\frac{1}{2}\sqrt{1-2\operatorname{D}^{2}_{ \phi, \phi'}}. \label{povm}
\end{eqnarray}
Apply this consideration to each element pure state in a mixed $\rho=\sum_{n}r_{n}\cdot|\phi_{n}\rangle\langle \phi_{n}|$, we  obtain
    \begin{eqnarray*}
\operatorname{G}_{A|\rho}&\leq& \textstyle\frac{1}{2}+\textstyle \max_{\{r_{n},\phi_{n}\}}\textstyle\sum_{n} r_{n}\cdot\textstyle\frac{1}{2}\textstyle\sqrt{1- 2\operatorname{D}^{2}_{\phi_{n}, \phi'_{n}}}\\
&\leq& \textstyle\frac{1}{2}+\max_{\{r_{n},\phi_{n}\}}\textstyle\frac{1}{2}\sqrt{1-2 \sum_{n} r_{n}\textstyle\cdot \operatorname{D}_{\phi_{n}, \phi'_{n}}^{2}} \\
&\leq& \textstyle\frac{1}{2}+\max_{\{r_{n},\phi_{n}\}}\textstyle\frac{1}{2}\sqrt{1- 2 (\textstyle\sum_{n} r_{n} \cdot  \operatorname{D}_{\phi_{n}, \phi'_{n}})^{2}} \\
&\leq&  \textstyle\frac{1}{2}+\textstyle\frac{1}{2}\sqrt{1-2\operatorname{D}^{2}_{\rho, \rho'}} \\
&\leq&  \textstyle\frac{1}{2}+\textstyle\frac{1}{2}\sqrt{1-2\operatorname{D}^{2}_{A\to B, \rho}},
\end{eqnarray*}
where the first inequality is due to  Eq.~\eqref{povm} and the second one is due to the concavity  of square rooting  and the third is due to convexity of squaring and the fourth is due to the convexity of $\operatorname{D}_{\rho,\rho}$ and   the last inequality is due to data processing inequality.     Thus,  we have shown that the disturbance effect, in state or in measurement, implies an upper-bound on the maximum correctly guessing probability, and
an operational lower bound of quantum randomness follows  as
 \begin{eqnarray}
\operatorname{H}^{C}_{min}(A|\rho)\geq -\log (\textstyle\frac{1}{2}+\textstyle\frac{1}{2}\sqrt{1- 2\operatorname{D}^{2}_{A\to B, \rho}} ). \label{rand}
\end{eqnarray}

\subsection{Proof of $\frac{1}{2}\delta_{\mathcal{A}:\mathcal{B}}\delta_{\mathcal{A}; \rho}\geq {\rm D}^{\varepsilon, \epsilon}_{\rm A\to B, \rho}$. }
We have the disturbance from a real measurement as ${\rm D}^{(re)}_{A\rightarrow B, \rho}:=\frac{1}{2}\sum_{i}|{\rm Tr}(\rho\cdot M^{(re)}_{b_i})-{\rm Tr}{[\Lambda_{re}(\rho)\cdot M^{(re)}_{b_i}]}|$ where $\Lambda_{re}$ denote the channel corresponding to real measurement $\mathcal{M}^{(re)}_{A}$.
 \begin{eqnarray}
 2 {\rm D}^{(re)}_{A\rightarrow B, \rho}&=&\sum_{i}\Big|{\rm Tr}(\rho\cdot M^{(re)}_{b_i})-{\rm Tr}(\rho\cdot|b_{i}\rangle\langle b_i|)+{\rm Tr}(\rho\cdot|b_{i}\rangle\langle b_i|)\nonumber \\
&&-{\rm Tr}(\Lambda_{re}(\rho)\cdot|b_{i}\rangle\langle b_i|)+ {\rm Tr}[\Lambda_{re}(\rho)\cdot|b_{i}\rangle\langle b_i|]-{\rm Tr}{[\Lambda_{re}(\rho)\cdot M^{(re)}_{b_i}]}\Big|\nonumber \\
&\leq& \sum_{i}\Big|{\rm Tr}(\rho\cdot M^{(re)}_{b_i})-{\rm Tr}(\rho\cdot|b_{i}\rangle\langle b_i|)|+\sum_{i}\Big|{\rm Tr}(\rho\cdot|b_{i}\rangle\langle b_i|)-{\rm Tr}(\Lambda_{re}(\rho)\cdot|b_{i}\rangle\langle b_i|)\Big|\nonumber \\
&&+ \sum_{i}\Big|{\rm Tr}(\Lambda_{re}(\rho)\cdot|b_{i}\rangle\langle b_i|)-{\rm Tr}{(\Lambda_{re}(\rho)\cdot M^{(re)}_{b_i})}\Big|\nonumber \\
&\leq & 4\varepsilon_{B} + \sum_{i}\Big|{\rm Tr}(\rho\cdot|b_{i}\rangle\langle b_i|)-{\rm Tr}(\Lambda_{re}(\rho)\cdot|b_{i}\rangle\langle b_i|)\Big|\nonumber\\
&\leq & 4\varepsilon_{B} + \sum_{i}\Big|{\rm Tr}(\rho'\cdot|b_{i}\rangle\langle b_i|)-{\rm Tr}(\Lambda_{re}(\rho)\cdot|b_{i}\rangle\langle b_i|)\Big|+\sum_{i}\Big|{\rm Tr}(\rho\cdot|b_{i}\rangle\langle b_i|)-{\rm Tr}(\rho'\cdot|b_{i}\rangle\langle b_i|)\Big|\nonumber \\
&\leq &  4\varepsilon_{B} +2\varepsilon_{A}+2{\rm D}_{{\rm A\to B}, \rho},
\end{eqnarray}
where according to the definition of $\varepsilon_{A}$ and  $\varepsilon_{B}$, we have   $\frac{1}{2}\sum_{i}\Big|{\rm Tr}(\varrho\cdot|b_{i}\rangle\langle b_i|)-{\rm Tr}{(\varrho\cdot M^{(re)}_{b_i})}\Big|\leq \varepsilon_{B}$  and  $\varepsilon_{A}\geq {\rm D}_{\rho, \Lambda_{re}(\rho)} \geq \frac{1}{2}\sum_{i}\Big|{\rm Tr}(\rho'\cdot|b_{i}\rangle\langle b_i|)-{\rm Tr}(\Lambda_{re}(\rho)\cdot|b_{i}\rangle\langle b_i|)\Big|$.
Then we have the lower bound on the state collapse as
 \begin{eqnarray}
  \frac{1}{2}\delta_{\mathcal{A};\mathcal{B}}\delta_{A, \rho}\geq  {\rm D}_{A\to B, \rho}\geq
 {\rm D}^{(re)}_{A\rightarrow B, \rho}-2\varepsilon_{B}-\varepsilon_{A}.
\end{eqnarray}
\subsection{Proof of Theorem~5}
With the definition of  $\operatorname{H}^{C}_{min,asy}(A|E)$, we have
\begin{eqnarray}
\operatorname{H}^{C}_{min,asy}(A|E)&=&\min_{r_{n}, \phi_{n}}\sum_{n} r_{n}\cdot \operatorname{H}(\mathbf p_{\phi_n}) \geq 2 \min_{r_{n}, \phi_{n}}\sum_{n}r_{n} \delta^{2}_{\mathcal{A};\phi_{n}} \nonumber \\
&\geq &4 \textstyle\sum_{n}r_{n}\tau^{2}_{A\to B, \phi_n}\geq  4(\textstyle\sum_{n}r_{n}\tau_{A\to B, \phi_n})^{2} \nonumber \\
 &\geq & 4\tau^{2}_{A\to B, \rho},
\label{rand4}
\end{eqnarray}
where  the second inequality is due to the inequality of binary entropy  $-p\log p-(1-p)\log (1-p)\geq 4p(1-p)$ and the  $\delta_{A;\rho}=\sqrt{2p(1|A;\phi)(1-p(1|A;\phi))}$ defined  in Eq.(\ref{defun}) and  the  third inequality is due to Eq.(\ref{g2}) and the fourth is due the convexity of $\tau_{A\to B, \rho}$.

\clearpage

\end{document}